\documentclass[amsmath,amssymb,aps,prl,reprint,longbibliography]{revtex4-1}

\usepackage{graphicx}% Include Fig.~files
\usepackage{color}

\usepackage{textcomp} %for text mu

\usepackage[colorlinks,linkcolor=blue,citecolor=blue,urlcolor=black]{hyperref}

\begin{document}

\title{Quantum remote sensing of angular rotation of structured objects}%

\author{Wuhong Zhang}
\author{Dongkai Zhang}
\author{Xiaodong Qiu}
\author{Lixiang Chen}
\email{chenlx@xmu.edu.cn}
\affiliation{Department of Physics, Jiujiang Research Institute and Collaborative Innovation Center for Optoelectronic Semiconductors and Efficient Devices, Xiamen University, Xiamen 361005, China}

\date{\today}

\begin{abstract}
Based on two-photon entanglement, quantum remote sensing enables the measurement and detection to be done non-locally and remotely. However, little attention has been paid to implement a noncontact way to sense a real object¡¯s angular rotation, which is a key step towards the practical applications of precise measurements with entangled twisted photons. Here, we use photon pairs entangled in orbital angular momentum (OAM) to show that a real object's angular rotation can be measured non-locally. Our experiment reveals that the angular sensitivity of the object encoded with idler photons is proportional to the measured OAM values of signal photons. It suggests potential applications in developing a noncontact way for angle remote sensing of an object with customized measurement resolution. Moreover, this feature may provide potential application in sensing of some light-sensitive specimens when the entangled photon pairs, which have significantly different wavelengths, are used, such as one photon is infrared but the other one is visible.

\end{abstract}

\maketitle
\section{I. INTRODUCTION}
A major application of optics is remote sensing and probing of a variety of properties of matter. Many techniques, based on the well known photons degree of freedom such as polarization, energy, time, wavelength and phase, have been elucidated over the years to achieve that goal\cite{schanda2012physical}. Recent years, as a new degree of freedom of photon, orbital angular momentum (OAM) has raised broad interest \cite{Padgett2017}. Laguerre-Gaussian (LG) modes are commonly used eigenstate of OAM \cite{allen1992orbital} and have been well studied. In particularly, due to the characteristic helical phase of LG mode, it has shown an attractive application in encoding optical image\cite{torner2005digital,zhang2016encoding} and retrieving topographic information of an object \cite{molina2007probing,petrov2012characterization}. By analyzing the LG mode spectrum of scattered beam, some new sensing and probing applications, such as the nanometric displacement detection \cite{wang2006nanometric}, the detection of individual biomolecules \cite{kravets2013singular} and the discrimination of enantiomers in biological chemistry \cite{brullot2016resolving}, which were also proposed. More recently, by detecting the frequency shift of on-axis OAM components that are scattered from a spinning object, the angular speed can be deduced \cite{lavery2013detection}. Based on the rotational doppler effect, we also achieved the remote object's rotation speed detection in 120m free space \cite{wuhong2018free}. By measuring the phase differences between a light beam¡¯s constituent OAM modes, the remote sensing of an object¡¯s rotational orientation was also implemented \cite{Milione2017}. However, it is noted that those sensing methods of object's rotational angular were only demonstrated in classical area.

Relative to those classical methods, quantum remote sensing with OAM-entangled photon pairs offers an improved angular resolution that is amplified by large OAM values \cite{Jha2011Supersensitive}. By using entanglement of high orbital angular momenta, Robert {\it et al.} proposed the idea of OAM-increased angular resolution in remote sensing \cite{fickler2012quantum}. Omar {\it et al.} showed that by using weak measurements in the LG mode azimuthal degree of freedom, an effective amplification of the weak value with a factors of 100 was achieved \cite{Omar2014}. Benefiting by a novel liquid crystal device, a q-plate, which efficiently maps pure polarization states into hybrid SAM-OAM states and vice versa, Vincenzo {\it et al.} implemented a photonic gear for polarization ultra-sensitive noncontact angular measurements \cite{d2013photonic}. However, it is noted that only phase or polarization angular sensitivity has been demonstrated in previous study. For practical application, a real object usually contains both intensity and phase information. Although correlated spiral imaging theory \cite{simon2012two}, quantum digital spiral imaging of a pure phase object \cite{chen2014quantum} and spatial symmetry analysis of a real object \cite{patarroyo2013} have been proposed based on the OAM-entangled photon pairs, quantum remote sensing of rotation of a real structured object has not been demonstrated yet. The implementation of a noncontact way to measure the angular sensitivity of a real structured object is an important step towards practical quantum remote sensing.

 In this paper, by employing the high-dimensional spatial mode entanglement generated by spontaneous parametric down-conversion, we encode a rotation structured object with both amplitude and phase information into the full set of LG modes in idler arm. This consideration is important for constructing a high quality ghost image in signal arm since we utilized the full-field quantum correlations of spatially entangled photons \cite{PRL2012fullfield}. Then we use an ICCD camera and OAM analyzer to see and quantitatively measure the object's rotation angle respectively. Our experimental results find that the phase shift of signal photons is equal to both of the angular displacement of idler objects and measured OAM number of signal photons, which the formation might be considered as a quantum version of rotational doppler effect. Our work clearly shows that the noncontact measurement of angular rotation of an object has been successfully achieved.

\begin{figure}[t]
\centering
\includegraphics[width=\columnwidth]{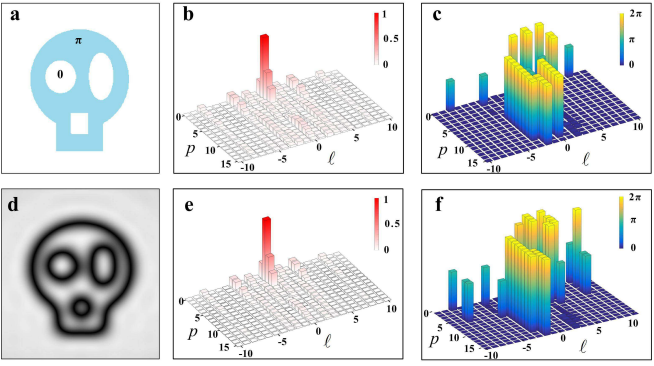}
\caption{Spectrum of a pure phase skull in LG mode bases. (a), pure phase skull object in idler arm, we define the phase of object is $\pi$, while the other part is $0$. (b), (c)the peak-normalized intensity spectrum, $|A_{\ell_i,p_i}|$ and phase spectrum, $arg(A_{\ell_i,p_i})$ of the skull object. Note that we have set the beam waist of the decomposing LG modes just equal to the size of the phase mask. This subtle consideration makes most of the constituent modes which are distributed over lower-orders. This is crucial for numerical reconstruction of the ghost image with a better fidelity, as two-photon OAM spectrum generated by SPDC is of limited spiral bandwidth, namely, lower OAM modes are produced more frequently than higher modes \cite{Torres2003quantum} (d) Reconstructed object intensity profile in signal arm, which can be regarded as a ghost image of the skull object, due to the destructive interference along the contours of a $\pi$ phase jump, the intensity of the ghost image can be outlined by the dark edges in the bright background. (e),(f) the peak-normalized intensity spectrum, $|B_{\ell_s,p_s}|$ and phase spectrum, $arg(B_{\ell_s,p_s})$ of the ghost image.}
\label{fig1}
\end{figure}
\section{II. THEORETICAL ANALYSIS AND SIMULATIONS}
The LG modes, $\rm{LG}_p^\ell(r,\phi )$ with azimuthal and radial indices $\ell$ and $p$, can offer a full basis to describe an arbitrary 2D transverse field in the high-dimensional Hilbert space \cite{simon2012two}. As the angle rotation of the object is only associated with the azimuthal indices $\ell$, LG modes provides a convenient basis for the measurement of rotation. At the single-photon level, the quantum state in the mode $\rm{LG}_p^\ell(r,\phi )$ can be labeled as $|{\ell,p}\rangle$. We assume that any structured object is described by a complex transmission function, $\psi(r,\phi)$. Then for a fundamental Gaussian illumination, the photon after this object acquires the complex amplitude $\psi(r,\phi)$, and its quantum state can then be denoted by $|\psi\rangle$. Based on the mode expansion method\cite{wuhong2018free,zhang2016encoding}, we have $|\psi\rangle=\sum_{\ell,p}{{A_{\ell ,p}}{|{\ell,p}\rangle}}$, where $A_{\ell,p}= \langle{\ell,p}|\psi\rangle$ denotes the overlap amplitude. As $A_{\ell,p}$ is a $(2\ell+1)\times(p+1)$ complex-valued spectrum, one need both the intensity spectrum, $|A_{\ell,p}|$, and phase spectrum, $arg(A_{\ell,p})$, to fully describe the object. Its significance also lies at the capacity to probe the phase objects, in addition to the amplitude ones. We show the numerical results of a pure phase skull in Fig.~\ref{fig1}, here, we restrict the decomposition with $\ell$ ranging from -10 to 10 and  $p$ from 0 to 15, totally involving 336 modes, which have reached nearly $98\%$ fidelity. Particularly, when the structured object is subject to a rotation by an angle $\alpha$, the intensity spectra remains unchanged but the phase spectra changed, due to a phase shift is induced for each constituent azimuthal mode, $\Delta\phi=\ell\alpha$, appearing as a product of the angular displacement $\alpha$ and the OAM number $\ell$. Subsequently, we can rewrite the transmitted or reflected photon's quantum state as:
\begin{eqnarray}
|\psi\rangle=\sum\limits_{\ell,p}{A_{\ell ,p}}{|{\ell,p}\rangle\exp(i\ell\alpha)},
\label{eq1}
\end{eqnarray}

Of particular interest is when considering the bi-photon high-dimensional spatial mode entanglement generated by spontaneous parametric down-conversion(SPDC). If the idler photon is subject to a state as described in Eq.\ref{eq1}, then the signal photon will collapse into the following state,
\begin{eqnarray}
|\varphi\rangle_s=\langle{\psi|\Psi\rangle_{\rm{SPDC}}}=\sum\limits_{\ell_s,p_s}{B_{\ell_s,p_s}}{|{\ell_s,p_s}\rangle\exp(i\ell_s\alpha)},
\label{eq2}
\end{eqnarray}
where $B_{\ell_s,p_s}=A^{*}_{\ell_i,p_i}C_{p_s,p_i}^{\ell_s,\ell_i}$ denotes the spectrum of "ghost" image of the object in the LG mode bases. $|\Psi\rangle_{\rm{SPDC}}$ is the complete spatial structure of entangled photons which can be described in the full LG modes as $|\Psi\rangle_{\rm{SPDC}}=\sum_{\ell_s,\ell_i,p_s,p_i}{C_{p_s,p_i}^{\ell_s,\ell_i}}|\ell_s,p_s\rangle|\ell_i,p_i\rangle $ \cite{Miatto2011full,Schneeloch2016}. $C_{p_s,p_i}^{\ell_s,\ell_i}$ denotes the amplitude probability of finding one photon in the signal mode $|\ell_s,p_s\rangle$ and the other in the idler mode $|\ell_i,p_i\rangle$. Then we can obtain the ghost image's intensity spectra $|B_{\ell_s,p_s}|$ and phase spectra $arg(B_{\ell_s,p_s})$ if we consider the rotation angle $\alpha=0$, as illustrated in Figs.~\ref{fig1}(e) and (f). One can even reconstruct the object intensity profile by summing all of the involved LG modes, as shown in Figs.~\ref{fig1}(d). Despite that the object is a pure phase one, the ghost image can still be outlined by the dark edges in the bright background, as a result of destructive interference along the contours of a $\pi$ phase jump. Moreover, Eq.~\ref{eq2} also provides us an intuitive way to understand that the bi-photon high-dimensional spatial mode entanglement can be used to perform ghost imaging.

\begin{figure*}[t]
\centerline{\includegraphics[width=1.2\columnwidth]{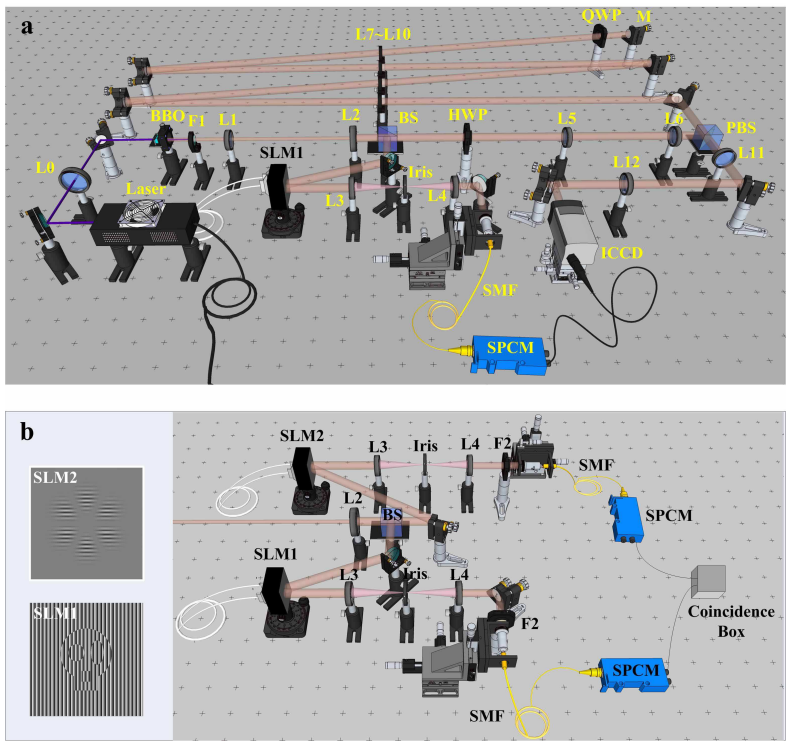}}
\caption{Schematic illustration of experimental setup. In the idler (object) arm of both setups, we use a 4f ($L_1$=100mm, $L_2$=300mm) system to image the crystal onto the spatial light modulator (SLM1) which displayed the rotatable object. Then we employ anther 4f ($L_3$=150mm, $L_4$=100mm) system to image the object onto the facet of the single-mode fiber, which is connected to a single-photon detector. Before each detector, a filter ($F_2$@710nm, bandwidth 10nm) is used to make sure that only the down conversion photon can be collected. In the signal arm, for ghost imaging setup (a): a half wave plate (HWP) is firstly used to rotate the polarization of the signal photons into vertical. Then there are three successive 4f imaging systems ($L_5$=750mm , $L_6$=750mm), ($L_7$=1000mm, $L_8$=1000mm), ($L_9$=1000mm, $L_{10}$=1000mm) consisting of the free-space delay line. The polarizing beam splitter (PBS) reflects the vertical signal photons, and at the end of the imaging system, after the vertical signal photon transmits twice of an quarter wave plate (QWP@ $22.5^\circ$), the polarization becomes horizontal, and goes back to the PBS, the followed 4f system ($L_{11}$=1000mm, $L_{12}$=1000mm) just images the crystal plane onto an intensified CCD (ICCD) camera, which is employed to record the ghost image of signal photons;  while for coincidence counting setup (b): the setup in signal arm is exactly the same as the idle ones, and the only difference is the displayed holographic gratings as shown in the inset of (b).}
\label{fig2}
\end{figure*}
Besides, in contrast with the local phase shift appearing in Eq.~\ref{eq1}, we can say the phase shift in Eq.~\ref{eq2} is nonlocal, as it is induced by the rotation of the object in the idler arm while the signal photons were left untouched. That's to say, by measuring the signal photon's phase shift, one can deduce the rotation angle of the object in a noncontact way. This constitutes the key idea to implement a quantum version of angle remote sensing of an object. To measure the signal photon's phase shift, we can use holographic method \cite{lavery2013detection,wuhong2018free} to filter out specific OAM modes superpositions such as ${|\psi \rangle_s}=(|\ell,p = 0\rangle+|-\ell,p = 0\rangle)/\sqrt 2$. Thus we can predict the bi-photon coincidence rate from Eq.~\ref{eq2} as,
\begin{eqnarray}
P_\ell(\alpha)=a_1^2+a_2^2+2a_1a_2\cos[2\ell\alpha+(\theta_1-\theta_2)],
\label{eq3}
\end{eqnarray}
where we have denoted the two complex amplitude $\sum_pA_{-\ell,p}^*C_{p_s= 0,p}^{\ell, -\ell}$ and $\sum_pA_{\ell,p}^*C_{p_s=0,p}^{-\ell,\ell}$ as $a_1e^{i\theta_1}$ and $a_2e^{i\theta_2}$, respectively. From Eq.~\ref{eq3}, a cosine curve will be observed in bi-photon coincidence rate and the period is determined by the angular displacement $\alpha$ in idler photon and the measured OAM number $\ell$ in signal photon. Moreover, interestingly enough, the rotation of the structured object could be non-locally detected with an angular sensitivity that is proportional by a factor $\ell$. It is noted that the decomposition spectrum of the structured object has some characteristic LG modes that are associated with its geometry. In our real experiment, we just utilize these dominant OAM modes for sensitive angle detection. Owing to the completeness and orthogonality, an arbitrary object can still be represented mathematically by an OAM spectrum in principle, such that it can be detected by our method with suitable OAM mode filtering.
\section{III. EXPERIMENTAL SETUP}
We build both the ghost imaging and coincidence counting setups and combine them to verify experimentally our theoretical predictions. In both setups, we use a 355nm ultraviolet laser to pump the type-I BBO crystal to generate bi-photon high-dimensional spatial mode entangled source collinearly, as shown in Fig. \ref{fig2}. We use a filter (F1) to block the 355 laser and, the idler and signal photons are then separated by a non-polarizing beam splitter (BS). In idler arm, the rotatable object is programmed into a suitable holographic gratings and then displayed by a spatial light modulator (SLM1). A single-mode fiber (SMF) connected with a single-photon detector (SPCM) is subsequently used to record the single-photon event, which does not have any spatial resolution. That's to say, there is no way to reconstruct the object only by recording the count of idler photon. The differences between two setups lie at the signal arms. In ghost imaging setup, as shown in Fig. \ref{fig2}(a), an intensified CCD (ICCD) camera is employed to record the ghost image, which is triggered by the single-photon events conveyed from the single-photon detector in the idler arm. Note that an optical delay of about 22 meters is requested to compensate the electronic delay \cite{morris2015imaging}. While in coincidence counting setup, as shown in Fig. \ref{fig2}(b), the object arm remains the same but there is no optical delay in the signal arm, instead, another SLM2 is employed to display specific holographic gratings to measure various OAM superposition states of signal photons. The setup in signal arm is exactly the same as the idler ones, and the only difference is the displayed holographic gratings as shown in the inset of Fig. \ref{fig2}(b). The outputs of the detectors are finally fed to a coincidence counting circuit with a time window of 25 ns.
\section{IV. EXPERIMENTAL RESULTS}
\begin{figure}[t]
\centerline{\includegraphics[width=\columnwidth]{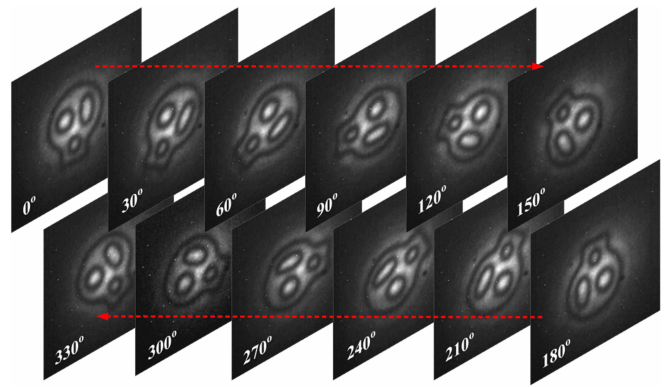}}
\caption{Experimental observations of ghost image as the skull object was rotated at $\alpha=30^\circ$ interval. Each ghost image was recorded by accumulating 1500 frames with 2s exposure time.}
\label{fig3}
\end{figure}

We first use the setup (a) of Fig.~\ref{fig2} to intuitively see the ghost image's rotation effect. We display the rotatable pure phase object of a skull on SLM1 in idler arm and record its ghost image with ICCD camera in signal arm. As we project idler photon into the state described in Eq.~\ref{eq1} with $\triangle\alpha=30^\circ$ each time, based on our theory, the signal photon will collapse into a state shown in Eq.~\ref{eq2}, namely, all of the involved LG modes will reconstruct the object in signal arm and its intensity is subsequently recorded by the ICCD. The experimental observation is shown by Fig. ~\ref{fig3}, as we rotate the skull object in idler arm, the ghost image in signal arm also rotates with the same angle, from which the excellent agreement with the theory can be seen. Here, each ghost image was recorded by accumulating 1500 frames with 2s exposure time. Moreover, each of the ghost image can be outlined by the dark edges as a result of destructive interference along the contours of a $\pi$ phase jump of the pure phase skull object in idler arm, which also show a good agreement with the simulation results in Fig.~\ref{fig1}(d). Here, we attribute the good quality of the ghost image to utilizing the full-field quantum correlations of spatially entangled photons which is using LG modes with both radial and azimuthal indices.

\begin{figure}[t]
\centerline{\includegraphics[width=\columnwidth]{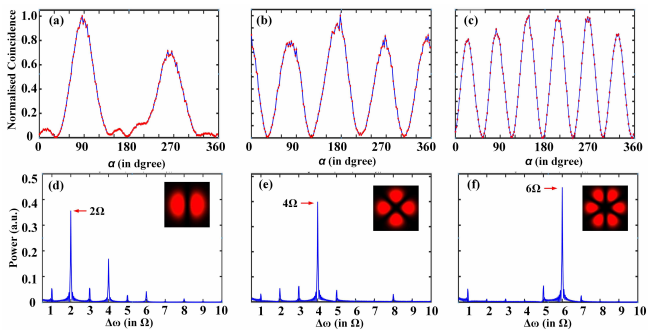}}
\caption{Experiment results for measuring the phase shift of the pure phase object in signal arm. Top line is the peak-normalized coincidence curves varying with the rotation angle $\triangle\alpha=2^\circ$, each data point was accumulated with 1 minute and averaged 5 times. Lower line is the beat frequencies by making a fast Fourier-transform (FFT) analysis of the corresponding coincidence curves, which shows an intuitive relation between the OAM numbers and the rotation speed $\Omega$. Here we define the rotation speed $\Omega=\alpha/t$.  The measured OAM superposition modes are $\ell=\pm1$ for (a) and (d), $\ell=\pm2$ for (b) and (e), $\ell=\pm3$ for (c) and (f). We also show the intensity profiles of the OAM superposition modes in the insets of (d), (e) and (f), respectively.}
\label{fig4}
\end{figure}

To quantitatively measuring the phase shift of the ghost image in signal arm, we further built a setup as shown in Fig.~\ref{fig2}(b) which is usually used to measure bi-photon high-dimensional spatial mode entanglement \cite{mair2001entanglement}. The skull object in idler arm now was driven to rotate at an interval of $\triangle\alpha=2^\circ$ each time, then we projected the signal photons to OAM superposition states of ${|\psi\rangle_s}=(|\ell,p=0\rangle+|-\ell,p = 0\rangle)/\sqrt 2$ and measured the two photon coincidence count. The recorded peak-normalized coincidence curves varying with the rotation angle $\alpha$ are illustrated in the top line of Fig.~\ref{fig4}. At each rotation angle, the accumulation time of coincidence was 1 minute and averaged 5 times. We measured three different OAM superposition states with $\ell$ increasing from 1 to 3 in signal arm, as shown in Figs.~\ref{fig4}(a)-(c). One can clearly see that the coincidence curves exhibit a cosine-like dependence on the rotation angle $\alpha$, which implies the phases of signal photons that are still shifted regardless of no rotation performed on the signal photons themselves. To show the cosine-like coincidence count signal deeply, we further performed a FFT of the signal with the rotation angle $\alpha$, as shown in Figs. ~\ref{fig4}(d)-(f). It is found that one-round rotation of the objects in the idler arm, which produces the coincidence counts with two-fold, four-fold and six-fold modulation rates in signal arm for $\ell=\pm1,\pm2,\pm3$, respectively. The reason underlying is because the $\pm\ell$  OAM components were phase shifted up and down, i.e., $\triangle\phi=\pm\ell\alpha$, resulting in a $2|\ell|\Omega$  modulation rate, which therefore confirms the prediction by Eq.~\ref{eq3}. It is noted that under the same phase shift, the angle sensitivity of the rotating object in idler photon is proportional to the measured OAM number in signal photon, which reveals that the object's rotation angular resolution is non-locally determined by a factor of $\ell$. This may have potential applications in developing a noncontact way for remotely sensing an object's rotation angle with customized resolution.

\begin{figure}[t]
\centerline{\includegraphics[width=\columnwidth]{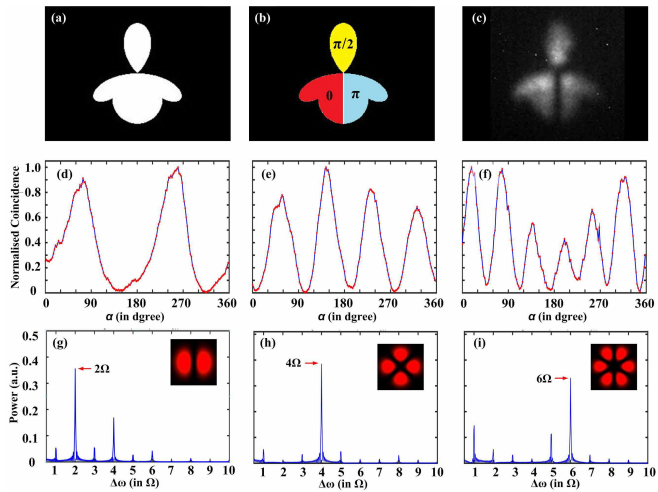}}
\caption{Experimental results for a real object with both intensity and phase information.} (a) and (b) are its intensity and phase profiles, respectively, while (c) is the recorded ghost image. (d)-(f) are peak-normalized coincidence curves by measuring OAM mode superpositions of $\ell=\pm1,\pm2,\pm3$, respectively. Here, in each rotation angle, the accumulation time of coincidence measurement was 5s and averaged 5 times. (g)-(i) show the angular frequency after a fast Fourier-transform analysis of the signal showed in (d), (e) and (f) with rotation angle $\alpha$, respectively. Here the rotation speed $\Omega=\alpha/t$.
\label{fig5}
\end{figure}

As we mentioned previously, for practical application, a real object usually contains both intensity and phase information, so we further exploit a complex-amplitude object, whose intensity and phase profiles are shown by Figs. \ref{fig5}(a) and (b), respectively. The structured object is designed to have a uniform intensity but with three different phases $0, \pi/2, \pi$ and there is a connection line between $0$ and $\pi$ area. Following the similar line but without the presence of the LG spectrum of the object, we first present the experimental ghost image of the object based on the setup of Fig.\ref{fig2}(a), as shown in Fig.\ref{fig5}(c). The ghost image's intensity profile of the object shows an additional dark line along the edge of $\pi$ phase jump compared with the intensity profiles of Fig.\ref{fig5}(a), which is exactly in accordance with the design. The good quality of the ghost image has demonstrated even further that the importance of using full-field quantum correlations of spatially entangled photons, such as both the radial and azimuthal indices of the LG modes. Similarly, as the object was rotated in the idler arm at an angular interval of $\alpha=2^\circ$, we measured the OAM superpositions of $\ell=\pm1,\pm2,\pm3$, and recorded the coincidence counts accordingly. The cosine-like varying curves with the rotation angle can be clearly seen in Figs.\ref{fig5}(d)-(f), which show again the signature of the nonlocal OAM-dependence phase shifts, $\triangle\phi=\ell\alpha$. Then, we further performed a FFT of the signal with the rotation angle $\alpha$ and see a more clearly highest angular frequency correspond two-fold, four-fold and six-fold of the rotation rates, as shown in Figs.\ref{fig5}(g)-(i). It is clearly shown that our proposed method can also be efficiently used for probing a real object with the complex-amplitude towards practical application of quantum remote sensing.

This method of remote sensing has a nonlocal feature because the signal photons used for sensitivity-increased angular detection do not illuminate the object in the idler arm. Moreover, this feature could be understood in a more intuitive way in light of ghost imaging. We can imagine that when the object is subjected to a rotation, its ghost image will be subjected to a rotation synchronously. Thus, regardless the signal photons do not touch the object, they can be considered to have traversed a virtual skull object, i.e., its ghost image. Thus, we can conclude that measuring the signal photons that interact virtually with the ghost image, which would just has a physical equivalence to that interact with the object, as manifested by Eq.~\ref{eq2}. Compared with previous ghost imaging system, which only the stationary objects are considered, here we use a rotation object which is challenging to obtain its ghost imaging in real time because of the extremely low photon-flux. However, it is noted that we only consider specific OAM components, without the need of ghost imaging of the full details of the object, to deduce the rotation angle of the object in a nonlocal and high-efficient way.

Also, it is illuminating for us to connect our quantitative measurement of the phase shifts to the rotational Doppler effect \cite{lavery2013detection,courtial1998measurement,courtial1998rotational}, namely, $\Delta\omega=\ell\Omega$, where $\Omega$ is the angular rate of the rotating object and $\ell$ is the detected OAM number. For this, we are allowed to replace the angular coordinate $\alpha$ in both Figs. \ref{fig4} and \ref{fig5} with the time coordinate equivalently, namely, $t=\alpha/\Omega$. In this scenario, a fast Fourier-transform (FFT) analysis of the coincidence counts is able to reveal a dominant peak at $2\Omega, 4\Omega$ and $6\Omega$ for $\ell=1, 2, 3$, respectively, as we have shown in both bottom panels of Figs. \ref{fig4} and \ref{fig5}. These beat frequencies are exactly equal to $2|\ell|$ times the rotation rate of the objects, which can also be inferred straight forwardly from Eq.~\ref{eq3} by simply replacing $\alpha$ with $\Omega t$. In contrast to previous classical experiments, the phase or frequency shifts are induced locally by a classical light beam that interacts with the rotating object, the beat signals here reveal some quantum aspects of the Doppler effect, as the phase shift of signal photons result nonlocally from the rotation of the objects placed in the idler arm that the signal photons do not illuminate.
\section{V. CONCLUSION}
In conclusion, by employing the full-field quantum correlations of spatially entangled photons which fully explored both radial and azimuthal indexes ($\ell, p$) of LG modes to describe various structured objects, we can not only obtain a high quality ghost image but also sense the object rotation angle in a nonlocal manner. And we are able to use one photon to illuminate the rotating object while use the other photon to detect the phase shift and measure the angular displacement, and thus demonstrating a new sensing application where the measured photons does not necessarily face or touch the object. Moreover, we have found that the phase shift of signal photons is equal to both of the angular displacement of idler objects and measured OAM number of signal photons. And the angular sensitivity of idler object is proportional to the measured OAM values of signal photons, which may have potential applications in developing a noncontact way for remotely sensing an object's rotation with customized resolution. As far as we know, this is the first experimental demonstration of a real structured object's rotation angle measurement in a nonlocal manner. Besides, we also have revealed that the noncontact rotation angle remote sensing scheme could be explained as a quantum version of rotational Doppler effect, with the amount of the nonlocal rotational frequency shift appearing as a product of the angular speed of the idler objects and the OAM number of the signal photons. The real frequency shift of signal photon in quantum version of rotational Doppler effect may be implemented with some bright bi-photon sources \cite{kuklewicz2004high,jeong2016bright}. Our present work may have the potential to develop a security technology for remote sensing also in military regime, where the detected light does not necessarily touch or face the dangerous rotating objects. The ability of our scheme working at the photon-count level may also suggest potential applications, e.g., in bioimaging and biosensing, in which a low-photon flux is essential as a high-photon flux might have detrimental effects \cite{morris2015imaging}. The natural extension of our scheme is to implement the free-space link for long-distance entanglement-enhanced remote sensing technique \cite{tamburini2011twisting,krenn2015twisted,krenn2016twisted,sit2017high,farias2015resilience}. In addition, by utilizing the correlated photons which have significantly different wavelengths, such as using the infrared photons at 1550 nm wavelength to illuminate object, but detecting with visible photons at a wavelength of 460 nm \cite{Aspden2015optica}, it may provide potential application in rotation sensing of some light-sensitive specimens, such as pure phase or complex amplitude organism in biological sensing.
\section{ACKNOWLEDGMENTS}
We are grateful to the Optics group led by Prof. Miles Padgett at the University of Glasgow for kind help in LabVIEW codes, and thank Mr. Ming Su at the University of Queensland, Dr. Robert Fickler, Enno Geise, Yingwen Zhang, Ebrahim Karimi at the University of Ottawa for useful discussions. This work is supported by the National Natural Science Foundation of China (91636109), the Fundamental Research Funds for the Central Universities at Xiamen University (20720190054,20720190057), the Natural Science Foundation of Fujian Province of China for Distinguished Young Scientists (2015J06002), and the program for New Century Excellent Talents in University of China (NCET-13-0495).

%\bibliographystyle{apsrev4-1}
%\bibliography{myref}

%

\end{document}